\documentclass[journal]{IEEEtran}
\usepackage{amsmath}
\usepackage{lineno,hyperref}
\modulolinenumbers[5]
\usepackage{amssymb}
\usepackage{booktabs}
\usepackage{graphicx}
\usepackage{bm}
\usepackage{cite}
\usepackage{multirow}
\usepackage{booktabs}
\usepackage[table,xcdraw]{xcolor}
\usepackage{pifont}
\usepackage{bbm}
\usepackage{float}
\begin{document}
	
\title{Implicit Structural Modeling via Generative Diffusion Frameworks}
\author{Yimin Dou, Xinming Wu*, Zhixiang Guo, Hui Gao, Buyu Deng	
		\thanks{The corresponding author is Xinming Wu}
	}
\maketitle

\begin{abstract}	
	Implicit structural modeling can support understanding subsurface spatial configurations, revealing patterns of geological evolution, and enabling quantitative simulation of geological processes, thereby offering substantial scientific and engineering value. Conventional approaches formulate it as an optimization problem or framework interpolation to fit a continuous scalar field, whereas machine learning methods typically adopt discriminative regression to directly predict implicit models. However, in complex scenarios involving fault intersections, branching, and thrust nappes, these methods still struggle to maintain topological consistency and kinematic plausibility. In this work, we develop an implicit structural modeling approach based on diffusion models. We construct a set of training data through a simulation based synthesis pipeline and design a dedicated encoder for conditional injection, allowing the conditional branch to converge rapidly while effectively reinforcing the input conditional priors throughout the diffusion process, thereby more stably propagating structural constraints. We then inject these conditional features into a backbone network pretrained on large scale natural images to enable conditional training of the diffusion model. Although our synthetic data include only a relatively stylized normal fault system, experiments demonstrate strong generalization, enabling the model to effectively handle diverse complex structural types such as strike slip faults and intricate flower fault systems. More importantly, even in challenging thrust nappe settings where the scalar field becomes non monotonic and exhibits abrupt depth discontinuities, the model can still generate reliable implicit structural models. 

	\end{abstract}
	\begin{IEEEkeywords}
	Implicit structural modeling, Diffusion model
	\end{IEEEkeywords}

\section{Introduction}

Structural models can support key engineering workflows such as stratal framework construction, trap identification, and reservoir modeling, while also providing continuous and computable geometric and topological representations for studies of structural evolution, sedimentary system reconstruction, and the mechanisms of geological hazards\cite{wellmann20183}. They therefore carry substantial engineering value and scientific significance. Conventional explicit modeling or surface based modeling approaches typically rely on geoscientists' interpretations of structures, digitizing the identified surface elements and their spatial extents to reconstruct complex structural geometries and spatial assemblages. The resulting models are often represented as a collection of geological interfaces and targets built through triangulation. However, these workflows are not only labor intensive but also highly dependent on subjective judgment, leading to poor reproducibility and limited repeatability\cite{caumon2009surface,chaodong2010rapid}.

Implicit modeling offers a more efficient, easily updatable, and reproducible alternative\cite{caumon2012three,hillier2014three,collon20153d}. It represents a given geological framework or target as a continuous scalar field and achieves a unified description of geological structures through operations such as iso-surface extraction. Because subsurface geological processes cannot be fully observed and structural data sampling is often heterogeneous, implicit modeling typically requires the introduction of prior geological principles as constraints, including smoothness, stratal continuity, and the plausibility of topological relationships. Compared with explicit models, implicit representations do not directly encode discrete semantic information such as stratal contacts and termination styles, and their ability to depict fine details is therefore relatively limited. Their advantages lie in a higher degree of automation, reduced dependence on input interpretation, and the natural production of structured field data that are readily represented on regular grids, making them inherently compatible with end to end learning and differentiable computation paradigms in AI \cite{bi2022deepismnet}.

At present, implicit structural modeling remains dominated by theory driven methods, with representative approaches including discrete smooth interpolation (DSI)\cite{mallet1988three,souche2014construction,renaudeau2019implicit}, potential field methods (PFM)\cite{lajaunie1997foliation,phillips2007use}, and their extensions. DSI formulates structural modeling as a least squares optimization problem with smoothness regularization on a volumetric mesh, thereby producing a continuous scalar field consistent with the input observations. PFM, often in a mesh free setting, reconstructs the scalar field directly from observations using interpolation strategies such as co-kriging or radial basis functions (RBFs). Although DSI and PFM differ in their implementation mechanisms, they share similar practical limitations. They tend to over rely on mathematical smoothness, approximating geological plausibility as a scalar field that is “as smooth as possible,” which makes it difficult to represent strong nonlinearity, discontinuities, and local abrupt changes inherent to structural processes. In addition, geological knowledge is typically translated into linear equality or inequality constraints, whose expressive power is limited, making it difficult to capture complex topological relationships such as fault intersections and branching, stratal truncation, unconformity incision, and thrust imbrication. Moreover, these methods are highly sensitive to data quality. When data are sparse, unevenly distributed, or affected by interpretation errors, they may yield models that appear geometrically reasonable but are geologically invalid.

DeepISMNet proposed by Bi et al. is the first implicit structural modeling framework that introduces deep learning and embeds geological constraints into a CNN loss function in a learnable manner \cite{bi2022deepismnet}. This approach partially replaces explicit mathematical constraints with data driven learning, and while maintaining structured outputs as continuous scalar fields, it can natively represent geological discontinuities, thereby enabling a more efficient, reproducible, and geologically consistent workflow. However, due to the intrinsic limitations of discriminative regression, it often learns only the dominant modes and average trends of the training distribution. It is therefore difficult to capture the inherent non-uniqueness and uncertainty of implicit modeling, to generate structures with more complex topologies and stronger nonlinear mechanisms, and to robustly transfer to novel structural types and assemblages that do not appear in the training data.

Generative models better match implicit structural modeling than discriminative ones. They can represent uncertainty from non-uniqueness and produce diverse structural styles and topologies under constraints, which improves out of distribution generalization and transfer to unseen structural types and complex systems. Typical frameworks include GANs \cite{goodfellow2014generative} and diffusion models \cite{ho2020denoising,rombach2022high}. From an inpainting perspective, implicit structural modeling interpolates horizons nonlinearly under fault skeleton constraints, and both GANs and diffusion models offer inpainting solutions.

Most GAN based inpainting extends regression by adding adversarial loss to reconstruction terms such as MSE \cite{pathak2016context,dou2023mda}. Because each condition usually has a single ground truth, training collapses the conditional variance, yielding near deterministic outputs that miss non-uniqueness. Pluralistic GANs introduce a samplable conditional prior via dual paths and distribution coupling \cite{zheng2019pluralistic}, but they often weaken instance level supervision on masked regions and rely more on adversarial constraints. For large missing areas or complex structures, this can cause blurred details, boundary inconsistencies, or semantic drift, trading accuracy and geometric consistency for diversity.

Diffusion inpainting follows two routes. RePaint style methods impose sampling constraints without extra training \cite{lugmayr2022repaint,mayettd} by repeatedly enforcing consistency in known regions during reverse sampling, using observed pixels or their forward noised versions at the current noise level. The other route trains conditional injection, feeding masks and known regions as extra channels or adding a conditional branch to propagate constraints throughout generation. RePaint style sampling is often unstable, sensitive to initial noise and trajectories, and may yield boundary artifacts or drift. In domain specific tasks, even with such constraints, substantial adaptation training is usually needed for reliable structural consistency. Conditional injection is generally more stable \cite{mallya2018piggyback,houlsby2019parameter,zhang2023adding,hu2022lora}: it injects priors such as skeletons through added modules while keeping the broadly pretrained backbone fixed, acting as a plug in domain knowledge module that supports transfer and generalization.

Backbone conditional injection is dominated by LoRA \cite{hu2022lora} and ControlNet \cite{zhang2023adding}. LoRA is lightweight but limited by low rank updates, which restricts expressive capacity and the bandwidth of injected signals, so it tends to work best within domain. For transferring from natural images to geoscience, the representation gap is large, and low rank perturbations often cannot supply the needed geometric and physical priors. ControlNet is more promising because it uses an independent conditional branch to propagate constraints, yet cross domain transfer remains challenging. Fault constraints are sparse and sharp, so alignment is easier. Horizon modeling demands stronger global consistency, including sequence and topology continuity, displacement relations, and principles such as monotonicity of relative geological time. Horizon signals are thin and weak, and can be overwhelmed by the backbone prior, causing breaks, drift, cross layer leakage, or failure to strictly satisfy constraints during sampling.

In this work, we propose a diffusion model framework for implicit structural modeling with dual conditional injection. Through a simple yet effective design, we introduce a dedicated encoder for conditional injection, enabling the conditional branch to converge rapidly while preserving and amplifying key structural signals during the diffusion process, thereby more stably propagating constraints from faults and horizons. By constructing a plug in geoscience knowledge base via an auxiliary conditional branch, the backbone network can acquire reliable structural prior injection without substantial modification for geoscience tasks, enabling effective generalization across broader structural types and geological settings, and improving robustness and transferability to out of distribution complex structural assemblages. The conditional branch serves solely as a structural information injection component, while the backbone can retain the general generative priors learned from large scale natural images. Consequently, even when training data are primarily derived from idealized simulation based synthesis, the model can still transfer to more complex real world structural settings and maintain stable performance and consistent modeling quality under out of distribution scenarios such as complex fault systems and thrust structures. These modeling results can not only be applied to a wide range of engineering and scientific studies, but also hold promise for building a data flywheel, in which more complex and realistic structural skeletons drive the implicit model to generate multimodal and paired consistent AI synthesized data, such as seismic, impedance, and gamma ray, gradually overcoming the stylized limitations of conventional simulation based synthesis and providing more realistic and diverse training samples for the training and evaluation of broader geoscience AI tasks.

\section{Approach}
\begin{figure*}[htb]
	\includegraphics[scale=0.62]{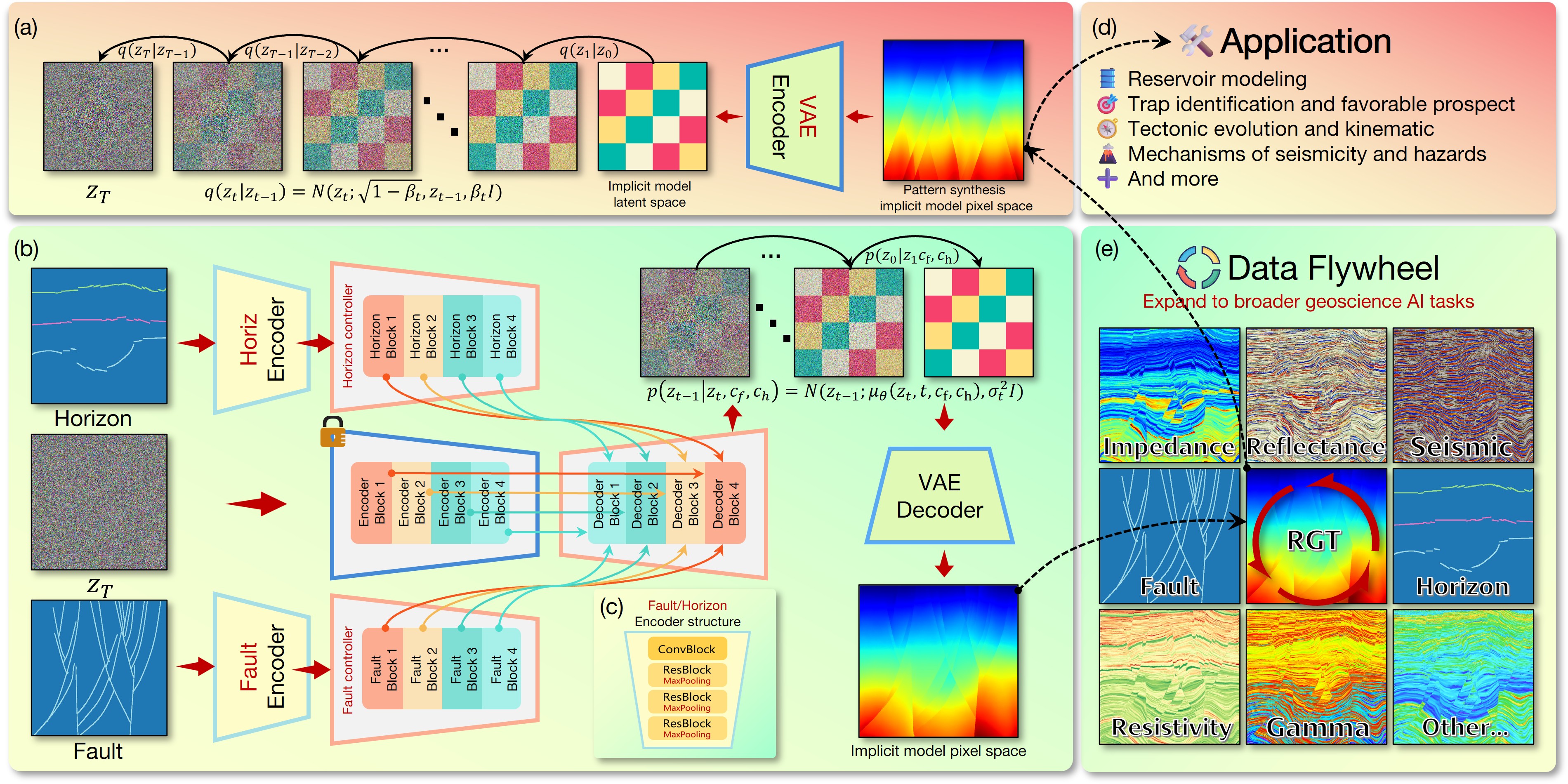}
	\centering\caption{(a) illustrates the forward noising process of the diffusion model. (b) presents the reverse sampling pipeline and the overall network architecture, including the definitions of inputs and outputs and the construction of the multi-branch conditional injection mechanism across the scales of the UNet decoder. (c) shows the architecture of the conditional encoder and its feature propagation path. (d) provides example results of the proposed method across multiple structural scenarios and representative applications. (e) illustrates the data flywheel enabled by this method, which can generate diverse data to support a broader range of Earth science AI tasks.}
	\label{fig1}
\end{figure*}
\subsection{Data preparation}
We follow a standard synthetic data generation pipeline and construct 4,000 paired 2D synthetic samples consisting of RGT fields and fault annotations, where the structural styles are dominated by stylized normal fault systems. During training, we extract sparse horizons and fault labels from the RGT field and use them jointly as conditional inputs to the diffusion model. Meanwhile, the complete RGT scalar field is used as the supervision signal to guide the model in learning implicit structural representations.

\subsection{Framework}

Figure \ref{fig1} (a, b) illustrates the overall framework of the proposed method. The diffusion component follows the Stable Diffusion  paradigm\cite{rombach2022high}, and the entire diffusion process is conducted in the latent space obtained via an 8× spatial compression by the VAE\cite{balle2018variational}. The forward process starts from the clean latent variable $z_0$ and progressively injects Gaussian noise according to a predefined noise schedule to obtain a sequence of $z_t$, until reaching $z_T$ that is close to an isotropic Gaussian distribution, which satisfies
\begin{equation}
\begin{aligned}
&q(z_t\mid z_{t-1})=\mathcal{N}\big(\sqrt{1-\beta_t},z_{t-1},\beta_t\mathbf{I}\big),\\
&q(z_t\mid z_0)=\mathcal{N}\big(\sqrt{\bar{\alpha}_t},z_0,(1-\bar{\alpha}_t)\mathbf{I}\big),
\end{aligned}
\end{equation}
where $\alpha_t = 1 - \beta_t$ and $\bar{\alpha}_t = \prod_{i=1}^{t} \alpha_i$. The reverse process starts from random noise $z_T$. Guided by the conditional inputs (fault labels and sparse horizon constraints), the denoising network iteratively predicts the noise (or equivalently the velocity or residual term) and progressively reconstructs $z_{t-1} \rightarrow z_0$. The conditional reverse transition can be written as
\begin{equation}
p_\theta(z_{t-1}\mid z_t, c_\text{f}, c_\text{h})=\mathcal{N}\big(z_{t-1}; \mu_\theta(z_t,t,c_\text{f}, c_\text{h}),\sigma_t^2\mathbf{I}\big),
\end{equation}
where,
\begin{equation}
\mu_\theta(z_t,t,c_\text{f}, c_\text{h})=\frac{1}{\sqrt{\alpha_t}}\left(z_t-\frac{\beta_t}{\sqrt{1-\bar{\alpha}_t}};\epsilon_\theta(z_t,t,c_\text{f}, c_\text{h})\right),
\end{equation}
$\epsilon_\theta(\cdot)$ denotes the conditional denoising network, and $c_\text{f}, c_\text{h}$ represents the conditional information. During training, an equivalent noise prediction objective is typically adopted.
\begin{equation}
\begin{aligned}\label{lossf}
&\mathcal{L}=\mathbb{E}_{z_0,t,\epsilon}\left[ \parallel \epsilon-\epsilon_\theta(z_t,t,c_\text{f}, c_\text{h}) \parallel_2^2\right], \\
&z_t=\sqrt{\bar{\alpha}_t}z_0+\sqrt{1-\bar{\alpha}_t}\epsilon,\ \epsilon\sim\mathcal{N}(0,\mathbf{I}).
\end{aligned}
\end{equation}
The resulting $z_0$ is then decoded by the VAE to produce the reconstructed implicit scalar field output.

\subsection{Conditional encoder}

The conditional encoder maps conditional information from pixel space to latent space, aligning it with $z_t$ in spatial scale. This facilitates the subsequent extraction of conditional features that can be injected into the backbone network, thereby influencing the reverse sampling trajectory of the diffusion model. Although one could directly use the VAE to encode the conditions, in practice the forms of conditions are often diverse. Structural constraints typically need to be combined with masks or require fine grained control over specific regions, whereas the input channels and interfaces of a VAE are usually fixed, making it difficult to flexibly satisfy these requirements. ControlNet commonly adopts a conditional branch composed of several consecutive purely convolutional modules, which is simple, efficient, and easy to customize, and is often sufficient for high frequency conditions such as edges, skeletons, or faults. However, for horizon conditions, which are relatively low frequency in the RGT pixel space yet carry explicit numerical constraints, a purely convolutional branch often has limited capability to preserve the constraints. The conditional information can be diluted by the strong generative prior of the backbone network, making it difficult to stably satisfy horizon constraints and to ensure global topological consistency of the geological structure.

We adopt a conditional injection strategy that better matches the requirements of geological structural modeling. Unlike control conditions in natural images, we aim for structural constraints to exhibit stronger “hard constraint” characteristics during learning and sampling, rather than merely achieving coarse agreement or allowing large diversity as soft control. To this end, in the downsampling stage of the conditional encoder, we use MaxPooling instead of stride 2 convolutions to preserve, as much as possible, the extrema and boundary responses of key structures such as faults and horizons. Meanwhile, we replace pure convolutional stacking with residual structures to continuously retain and propagate structural information during feature transformations, preventing the conditional signal from progressively attenuating in deeper layers. This process can be expressed as follows.
Let the conditional input be $c$, and let the output of the $l$-th layer of the conditional encoder be $h^{(l)}$, then,
\begin{equation}
\begin{aligned}
&h^{(0)} = \phi_0(c), \\
&\tilde{h}^{(l)} = \mathrm{MaxPool}\left(h^{(l)}\right),(l=0,\dots,L-1) \\
&h^{(l+1)} = \tilde{h}^{(l)} + \mathcal{F}_{\theta_l}\left(\tilde{h}^{(l)}\right),
\end{aligned}
\end{equation}
where $\phi_0(\cdot)$ is the initial mapping $3 \times 3$ convolution, and $\mathcal{F}_{\theta_l}(\cdot)$ denotes the residual branch. The final conditional feature $f_c = h^{(L)}$, aligned with $z_t$, is then obtained.  Figure \ref{fig1} (c) provides a graphical illustration of the architecture of the encoder.

\subsection{Conditional injection}

As discussed above, in conditional injection for diffusion models, high frequency skeletal information such as faults is usually the easiest to encode and stably inject, because it more closely resembles edge cues in natural images and does not carry explicit pixel value constraints. In contrast, horizon constraints are more low frequency in the RGT pixel space, contain continuous numerical information, and must strictly satisfy global geological topological consistency, making them harder to enforce effectively through conventional conditional branches. If faults and horizons are simply concatenated and injected through a single auxiliary branch, conditional competition can arise: fault features with strong boundaries and high responses tend to dominate the feature space, thereby overriding or diluting the weaker horizon signals and preventing the horizon constraints from being stably preserved. The root cause is that the two types of constraints have inherently different characteristics. Faults are closer to pure skeletal constraints, whereas horizons behave more like composite constraints of “skeleton plus numerical inpainting,” requiring both geometric continuity and numerical consistency to be maintained. Motivated by this, we construct two separate conditional branches to represent these two types of constraints, and integrate them into the backbone network at different levels and through different injection mechanisms. This enables decoupled modeling of faults and horizons and yields more robust conditional control.

We construct two conditional branches to represent different types of structural constraints. Each branch follows the same injection procedure. Let the condition be encoded into a latent space feature $f_c$ that is aligned with the latent space. The branch network consists of four scale modules, where the $k$-th module outputs a conditional feature $f_c^{(k)}$ for $(k = 1, \dots, 4)$. To control the injection strength, we apply a conditional scaling coefficient $s_k$, and then use a zero initialized convolution to project the conditional feature into the channel space consistent with the backbone feature. This makes the conditional injection zero at the beginning of training, thereby stabilizing the backbone behavior, and allows the model to gradually learn effective conditional residuals during training. Finally, the projected conditional residuals are injected via skip connections into the corresponding levels of the UNet decoder that match their spatial resolutions. This process can be formalized as follows,
\begin{equation}
	\begin{aligned}
f_c^{(0)}=\mathcal{E}(c), \ \
f_c^{(k)}=\mathcal{B}_k\left(f_c^{(k-1)}\right),\ \ k=1,2,3,4,
\end{aligned}
\end{equation}
where $\mathcal{E}(\cdot)$ denotes the conditional encoder and $\mathcal{B}_k(\cdot)$ denotes the $k$-th scale module. Introducing a zero-initialized convolution $\mathcal{Z}_k(\cdot)$, the conditional injection term at the $k$-th level is given by
\begin{equation}
\Delta^{(k)} = \mathcal{Z}_k\left(s_k, f_c^{(k)}\right),
\ \ \mathcal{Z}_k(\cdot)\ \text{zero-init},
\end{equation}
and is injected in a residual form at the corresponding decoder scale as follows,
\begin{equation}
\tilde{d}^{(k)} = d^{(k)} + \Delta^{(k)}.
\end{equation}
If the two conditional branches are $c_\text{f}$ and $c_\text{h}$, the fusion at the same scale can be written as
\begin{equation}
\tilde{d}^{(k)} = d^{(k)}
+ \mathcal{Z}_k^{(\text{f})}\left(s_k^{(\text{f})} f_{c_\text{f}}^{(k)}\right)
+ \mathcal{Z}_k^{(\text{h})}\left(s_k^{(\text{h})} f_{c_\text{h}}^{(k)}\right).
\end{equation}
where $d^{(k)}$ denotes the UNet decoder feature map at the $k$-th spatial resolution, and $\tilde{d}^{(k)}$ denotes the feature map after conditional injection. In this way, multi-scale conditional residuals can stably influence the reverse sampling process while avoiding disruption of the pretrained backbone's generative prior at the early stage of training.
\begin{figure*}[!htb]
	\includegraphics[scale=0.7]{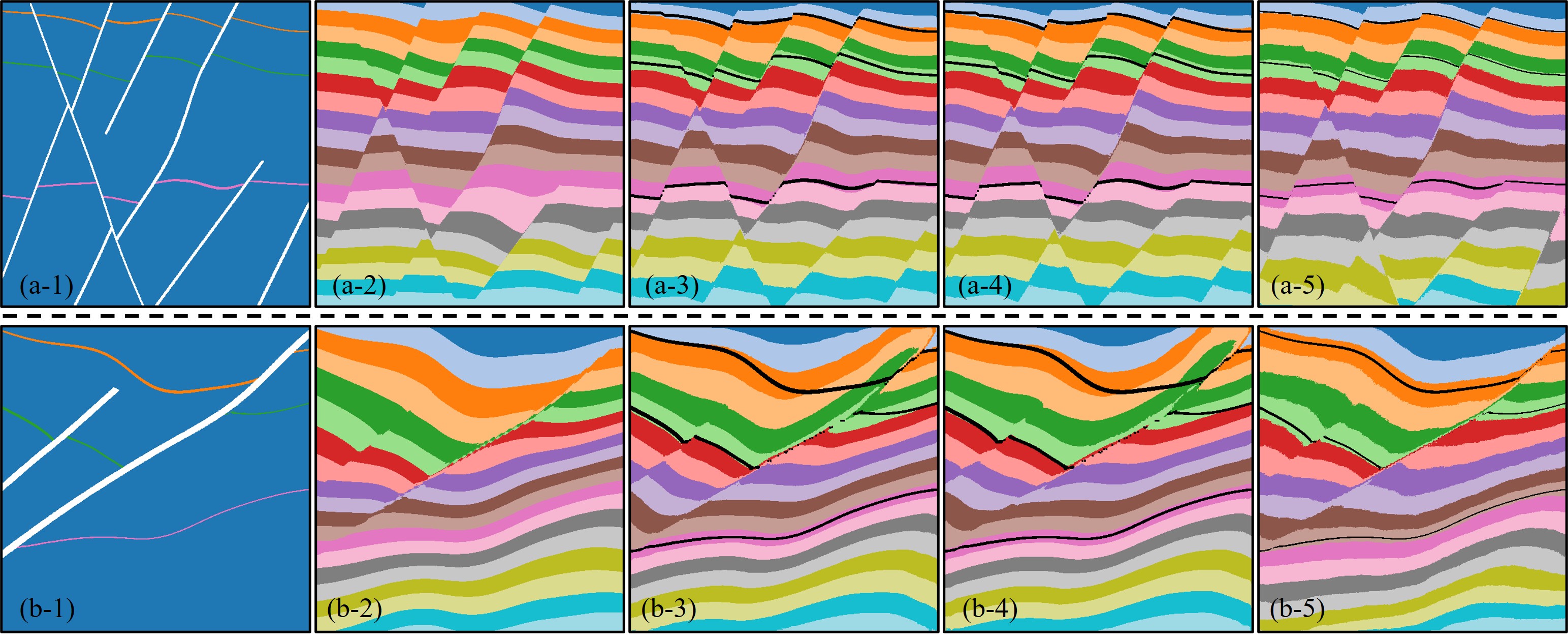}
	\centering\caption{Comparative results on the synthetic data. Panels (a-1), (b-1) show the conditional inputs to the model, where the white curves denote faults and the remaining colors represent horizons. Panels (a-2), (b-2) are the labels from the synthetic dataset, which correspond to only one possible solution given the same set of input constraints. Panels (a-3), (b-3) present the results of SD-UNet-Reg, panels (a-4), (b-4) show the results of SD-UNet-GAN, and panels (a-5), (b-5) are the results produced by our method. The black lines indicate the horizon constraints provided as inputs to the model.
	}
	\label{fig2}
\end{figure*}
\subsection{Training}
Because the UNet in SD has been extensively trained on large-scale natural image data, we aim to preserve the general generative priors it has learned as much as possible. We therefore freeze the encoder of the UNet to stably retain the backbone’s representational capacity. Meanwhile, given the substantial domain gap between natural images and RGT representations, we unfreeze the decoder so that, guided by the conditional branches, it can adapt to the geoscience setting and produce higher-quality implicit modeling results. In terms of the overall network architecture, the trainable components include the two controllers (conditional injection) and the decoder of the backbone.

We adopt the DDPM framework for training\cite{ho2020denoising}. During training, we start from the latent variable $z_0$ corresponding to a real sample. In the forward diffusion process, Gaussian noise is injected into $z_0$ according to the noise schedule to construct the noisy latent variable at an arbitrary time step $t$.
Subsequently, the conditional denoising network takes $(z_t, t, c_\text{f}, c_\text{h})$ as input and predicts the noise term $\epsilon_{\theta}(z_t, t,  c_\text{f}, c_\text{h})$ (or an equivalent velocity or residual parameterization) at each time step, using the noise prediction error as the training objective. This procedure is expressed in equation (\ref{lossf}).

During inference, we employ the deterministic (or low-stochasticity) sampling strategy of DDIM\cite{song2020denoising} to generate $z_0$ from $z_T$ progressively along the reverse trajectory, thereby obtaining stable implicit structural modeling results.

Training was completed on two NVIDIA H20 GPUs. We apply an exponential moving average (EMA) over the model parameters to improve training stability and enhance generation quality during inference. The batch size is set to 50, the learning rate is $1\times10^{-5}$, and AdamW is used as the optimizer, with a total of 200,000 training steps.

\section{Results}
\subsection{Experimental setup and data}

We primarily compare deep learning based approaches for implicit structural modeling. To ensure a fair comparison, all methods adopt the same backbone, namely the UNet architecture used in Stable Diffusion, with the diffusion time embedding removed. The methods differ only in their learning paradigms, including the discriminative regression framework proposed by Bi et al. and a generative GAN based framework, denoted as SD-UNet-Reg and SD-UNet-GAN, respectively. Because horizon constraints are highly sparse, plausible implicit modeling solutions are inherently non unique and diverse. Therefore, directly comparing per pixel or per patch absolute error metrics such as MSE or SSIM on the validation set is not sufficiently meaningful.

This chapter focuses on qualitative comparisons. We first present results on randomly synthesized samples, dominated by stylized normal fault systems, and then evaluate generalization to structural styles that are absent from the training set, including various flower fault systems, and compressional settings such as thrust or reverse fault systems. Finally, we test on several real world structural cases and highlight modeling results on complex structures from representative published studies.

\subsection{Synthetic data results}

\begin{figure*}[htb]
	\includegraphics[scale=0.49]{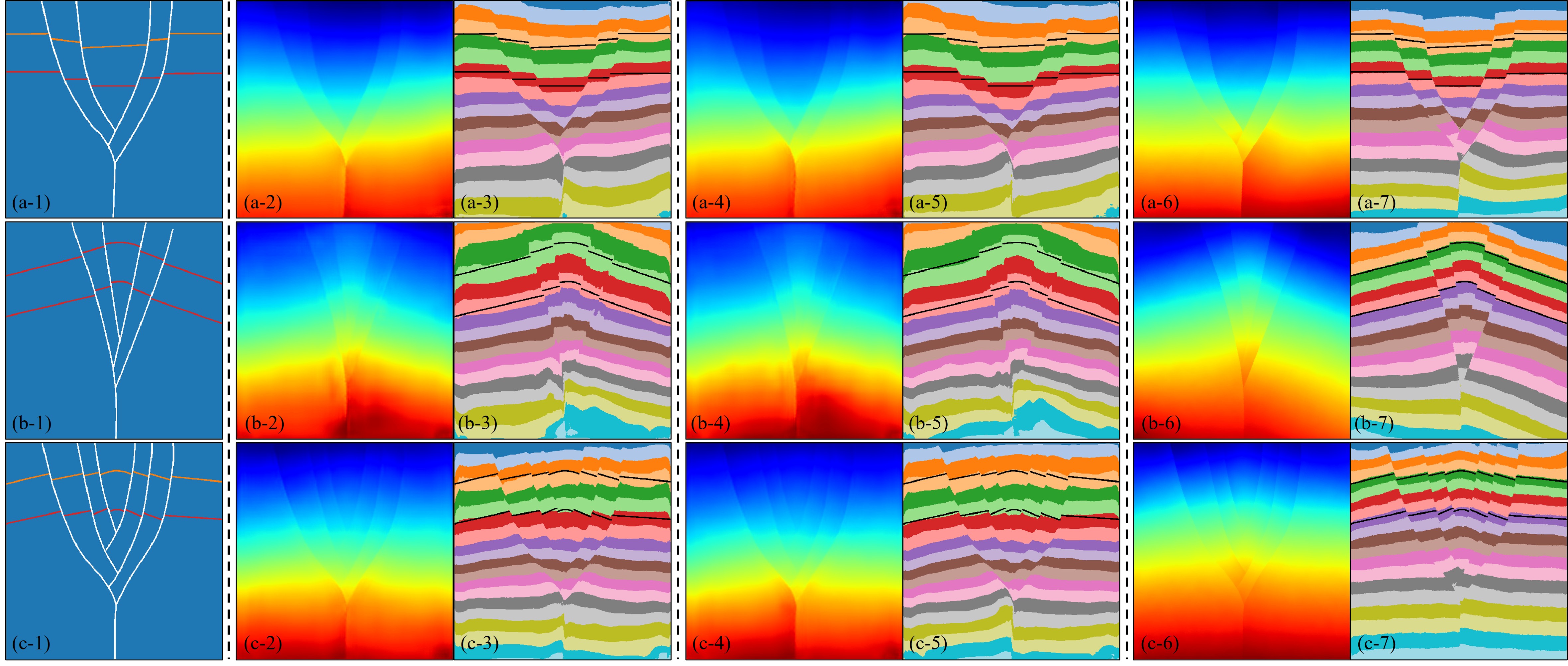}
	\centering\caption{Comparative results on strike-slip fault units. Panels (a-1), (b-1), and (c-1) show the conditional inputs to the model, where the white curves denote faults and the remaining colors represent horizons. Panels (a-2,3), (b-2,3), and (c-2,3) present the results of SD-UNet-Reg, panels (a-4,5), (b-4,5), and (c-4,5) show the results of SD-UNet-GAN, and panels (a-6,7), (b-6,7), and (c-6,7) are the results produced by our method. The black lines indicate the horizon constraints provided as inputs to the model.}
	\label{fig3}
\end{figure*}
We first compare the three methods on synthetic data. In addition, we generate an extra set of synthetic examples for qualitative evaluation. These samples are not used for training and are largely dominated by stylized normal fault systems. Figure \ref{fig2} provides qualitative comparisons on representative test cases. For relatively simple normal fault settings, both the SD-UNet-Reg and SD-UNet-GAN yield reasonable implicit modeling results (e.g., (a-2), (a-3)). However, when the fault geometry deviates noticeably from the training distribution, such as the steeply dipping fault in (b-1), both methods often exhibit poor constraint adherence and structural distortion (see (b-3) and (b-4)). In contrast, our method demonstrates better adaptability and stability across different structural scenarios. More importantly, it not only covers the distribution spanned by the training and test sets, but can also produce unexpected yet geologically plausible solutions under the given constraints, indicating stronger capability in representing non uniqueness and generalizing beyond the observed distribution.

To further examine the model’s adaptability to geometric combinations in strike slip settings, we additionally constructed a set of hand drawn strike slip flower fault units as conditional inputs. These designs cover synform, antiform, and hybrid branching styles, together with variations in branch number and dip configurations. In the absence of any ground truth scalar field constraints, this setting allows us to test whether the model can generate topologically consistent and geologically plausible implicit scalar fields under a more freely specified structural skeleton.

Following the taxonomy summarized by Huang et al. \cite{huang2017three}, we consider three idealized cross sectional models of flower structures in strike slip fault zones (Figure \ref{fig3} (a-1), (b-1), (c-1)). A negative flower structure exhibits a shallow synform, bounded by upward diverging strike slip fault splays, with normal displacement dominating along the branches. A positive flower structure shows a shallow antiform, likewise controlled by upward diverging splays, but with reverse displacement as the dominant component. Beyond these two classical end members, a hybrid flower structure also displays an antiform in the shallow section, while the branching faults are dominated by normal displacement; at depth, the splays converge into a near vertical master fault, representing a transitional strain state between the positive and negative flower structures.

Both SD-UNet-Reg and SD-UNet-GAN exhibit pronounced artifacts on strike-slip flower-fault cases. Typical failure modes include local numerical oscillations, spurious striping, or blocky noise near fault convergence and branching zones; horizons become unrealistically warped or broken when crossing faults; and, in some cases, cross-horizon “leakage” and non-conservative throw (i.e., inconsistent displacement relationships) are observed. These behaviors indicate that the two baselines have limited capability when confronted with complex strike-slip structures outside the training distribution. Although they can recover the overall trend of the implicit scalar field, they often produce low-quality local reconstructions in structurally critical regions (e.g., branching belts, intersection zones, and shallow flower units), making the horizon constraints difficult to satisfy consistently and ultimately degrading global topological consistency and geological interpretability.

As shown in Figure \ref{fig3} (a-6,7), (b-6,7), and (c-6,7), our method demonstrates clear advantages for strike-slip flower-fault scenarios. On the one hand, the generated implicit scalar field remains continuous and smooth in fault convergence and branching regions, without noticeable oscillations or artifacts such as striping or blocky noise. On the other hand, horizons preserve clear and geologically reasonable throw and geometry when crossing faults, satisfying the sparse horizon constraints while maintaining stratigraphic continuity and global topological consistency. Notably, even within shallow flower units and densely intersecting multi-splay zones, the model produces structurally stable and geologically interpretable results, indicating stronger out-of-distribution generalization and better adherence to complex structural constraints.

\begin{figure*}[htb]
	\includegraphics[scale=0.58]{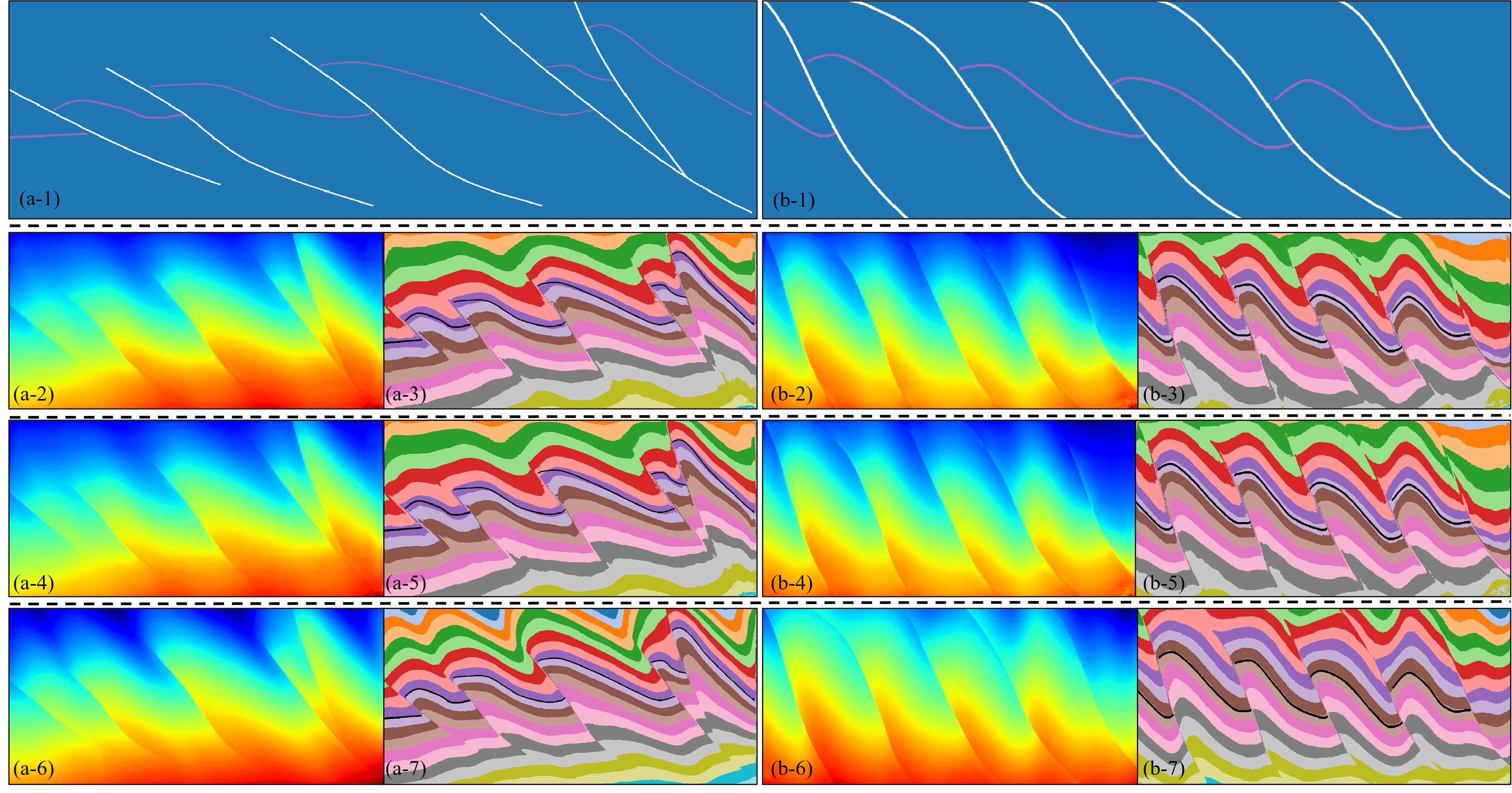}
	\centering\caption{Comparative results on thrust-fault units.
		 Panels (a-1) and (b-1) show the conditional inputs to the model, where the white curves denote faults and the remaining colors represent horizons. Panels (a-2,3) and (b-2,3) present the results of SD-UNet-Reg, panels (a-4,5) and (b-4,5) show the results of SD-UNet-GAN, and panels (a-6,7) and (b-6,7) are the results produced by our method. The black lines indicate the horizon constraints provided as inputs to the model.
}
	\label{fig4}
\end{figure*}

As shown in Figure \ref{fig4}, we constructed two simplified subduction–thrust examples to evaluate the model’s capability and robustness under strong compressional settings, particularly for non-monotonic horizons, deep structural jumps, and complex topological relationships. Overall, SD-UNet-GAN, SD-UNet-Reg, and our method can all produce a basic implicit model for these reverse-fault scenarios, but they differ markedly in fine-scale quality. SD-UNet-GAN and SD-UNet-Reg are more prone to local artifacts, such as stripe-like oscillations and blocky noise near fault neighborhoods or along data boundaries, as well as horizon breaks or excessive smoothing across displacement zones. In addition, some regions exhibit low-resolution blurry reconstructions and boundary sticking, making the horizon constraints difficult to satisfy consistently and degrading global topological consistency. In contrast, our method maintains a cleaner scalar field and sharper horizon geometry in structurally critical areas such as the fault vicinity and the thrust belt, leading to more stable and geologically consistent modeling results.

\subsection{Real data results}

\begin{figure*}[htb]
	\includegraphics[scale=0.52]{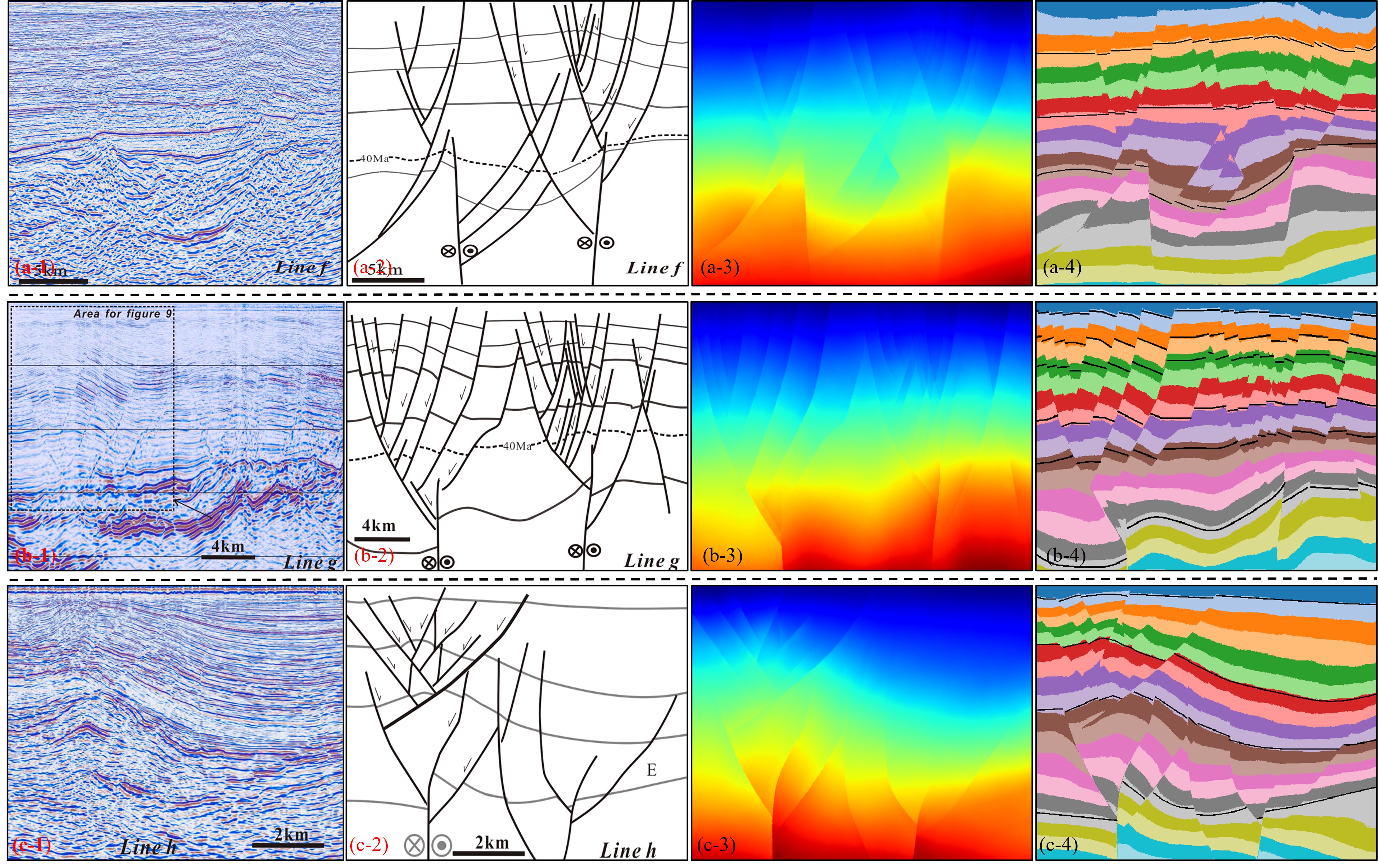}
	\centering\caption{Real-data modeling results for strike-slip flower-fault systems.
		Panels (a-1,2), (b-1,2), and (c-1,2) are reproduced from Huang et al.\cite{huang2017three} Panels (a-1), (b-1), and (c-1) show the original seismic sections of the three structural cases, while panels (a-2), (b-2), and (c-2) provide the conditional inputs to our model. Panels (a-3,4), (b-3,4), and (c-3,4) present our implicit structural modeling results for the three complex strike-slip flower-fault systems.The black lines indicate the horizon constraints provided as inputs to the model.}
	\label{fig5}
\end{figure*}

Next, we evaluate our method on several more complex real-world data, focusing on two structural styles that are notoriously challenging for traditional approaches: strike-slip flower-fault systems and thrust systems. We first present several representative strike-slip flower-structure examples discussed by Huang et al.\cite{huang2017three} In their work, the structures along these three seismic lines are all classified as hybrid flower structures. Specifically, Figure \ref{fig5} (a-1) and (b-1) are located within a right-lateral, left-step step-over zone, where a hybrid flower-structure unit develops on one side of each section; \ref{fig5} (c-1) lies within a restraining bend of the strike-slip fault system and contains two hybrid flower structures.

Figure \ref{fig5} (a-3,4), (b-3,4), and (c-3,4) show our modeling results on the above real datasets, demonstrating that the proposed method can stably satisfy both the fault-skeleton constraints and the sparse horizon constraints within complex strike-slip fault systems. Specifically, the reconstructed implicit scalar field remains continuous and clean in fault convergence and branching zones, without obvious artifacts, and produces clear and geologically reasonable throw across the faults. Meanwhile, the recovered horizons preserve stratigraphic continuity and global topological consistency. Even in highly deformed areas such as restraining bends and step-over zones, the method effectively avoids cross-horizon leakage and unrealistic inter-layer interpenetration.

For the structure corresponding to Figure \ref{fig5} (c-1), Huang et al.\cite{huang2017three} reported the coexistence of a doubly plunging anticline and multiple NE-striking normal faults, with the anticline axis oriented at approximately 45° to the main strike-slip fault; the shallow faults are dominated by normal throw and undergo slight upward rotation with depth. Our results better preserve this geometric coupling between folding and shallow normal faulting, enabling horizons to maintain consistent throw across faults while remaining globally topologically coherent, thereby providing a more stable and geologically consistent implicit modeling solution for hybrid flower structures.

\begin{figure*}[htb]
	\includegraphics[scale=0.526]{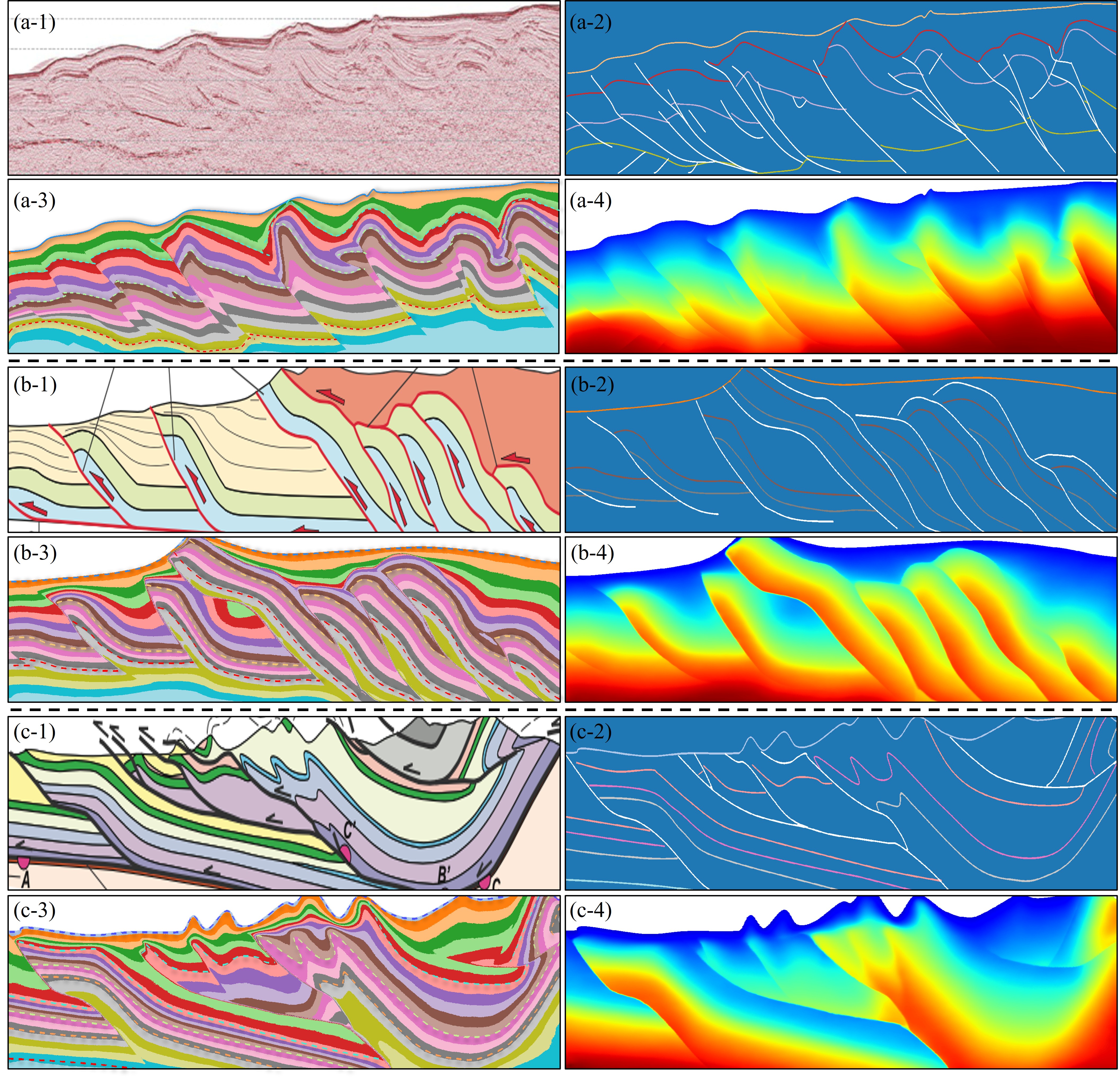}
	\centering\caption{Real-data modeling results for thrust–nappe systems.
		Panels (a-1), (b-1), and (c-1) are reproduced from Butler et al.\cite{butler2020thrust}, showing the seismic sections and/or interpreted models of three thrust-related structural cases. Panels (a-2), (b-2), and (c-2) present the conditional inputs to our model, where the white curves denote faults and the remaining colors represent horizons. Panels (a-3,4), (b-3,4), and (c-3,4) show our implicit modeling results for the three thrust-structure types, respectively. The dashed lines denote the horizon constraints provided as inputs to the model.}
	\label{fig6}
\end{figure*}

\begin{figure*}[htb]
	\includegraphics[scale=0.526]{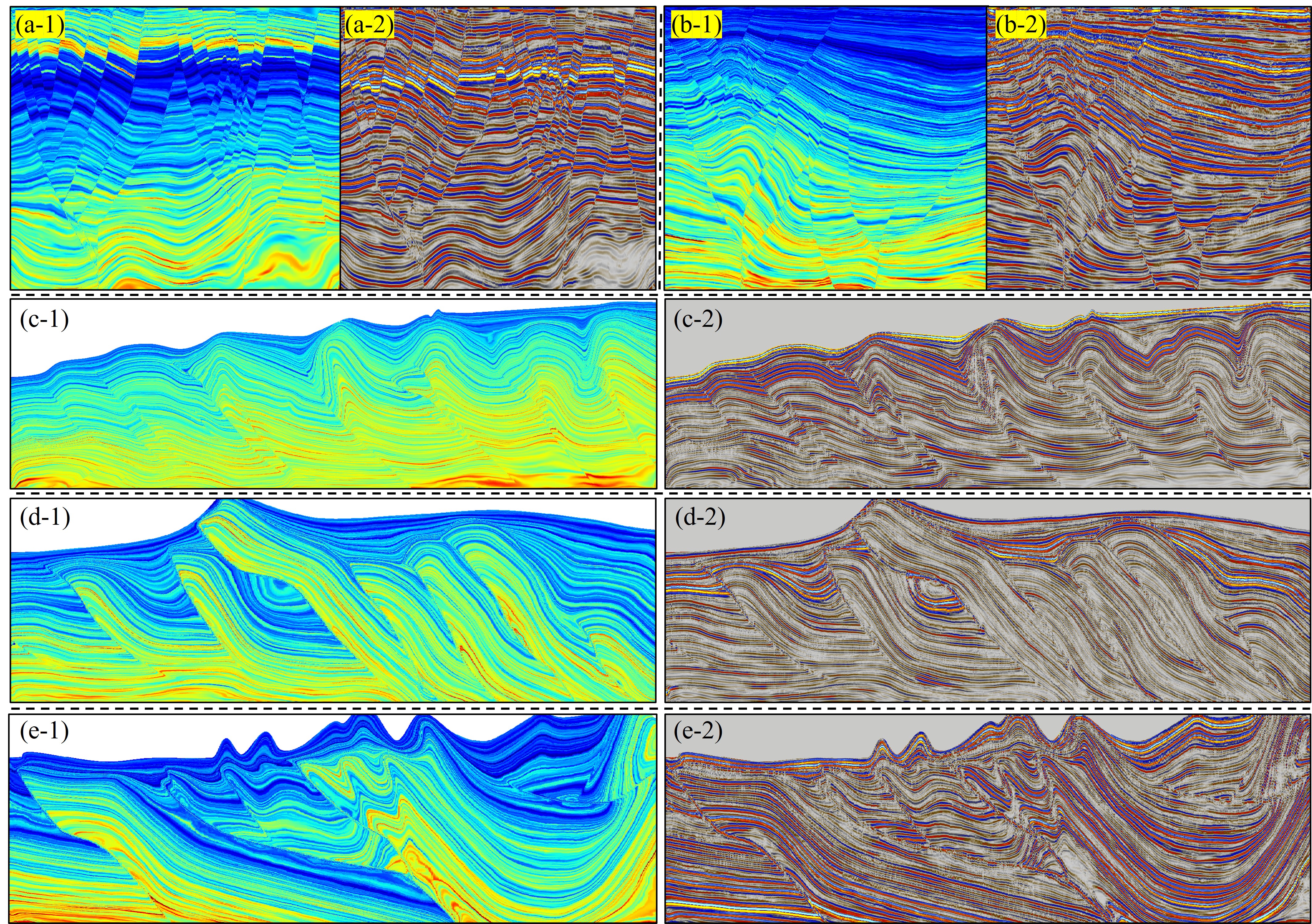}
	\centering\caption{Some examples of the data flywheel. (a-1), (c-1), (d-1), and (e-1) show the velocity models synthesized based on the results produced by our method, while (a-2), (c-2), (d-2), and (e-2) show the corresponding seismic data. These are illustrative examples. Following a similar workflow, we can synthesize additional properties and structural skeletons, thereby extending the approach to a broader range of geoscience and geophysical tasks.}
	\label{fig7}
\end{figure*}

Next, we select three complex thrust–nappe systems discussed by Butler et al.\cite{butler2020thrust} as our test benchmarks. These cases respectively cover (i) a submarine fold–thrust belt whose kinematic timing and migration are clearly recorded by growth strata (Figure \ref{fig6} (a-1)), (ii) an idealized fold-and-thrust system comprising both emergent and buried thrust geometries (Figure  \ref{fig6} (b-1)), and (iii) a thin-skinned thrust example characterized by a well-defined detachment layer and forward propagation of displacement (Figure \ref{fig6} (c-1)). Together, they allow us to systematically evaluate the robustness and generalization of our method under multi-scale structural complexity and distinct genetic constraints.

Figure \ref{fig6} (a-1) shows a deep-water submarine fold–thrust belt along the NW Borneo continental margin from a marine seismic section. It represents a gravity-collapse–driven, thin-skinned contractional system: the sedimentary wedge is detached along a weak, low-strength horizon (e.g., mud-rich units and/or evaporites), and the downslope displacement is accommodated at the toe of the slope by an imbricate array of thrusts and associated folds, paired with extensional deformation in the upslope shallow-water domain. In the interpreted section, the stratigraphy can be divided into pre-kinematic units, syn-kinematic growth strata, and post-kinematic cover.

Our modeling results are shown in Figure \ref{fig6} (a-3), (a-4). With only the fault skeleton and sparse horizon constraints as inputs, the model reconstructs an implicit scalar field consistent with the interpreted section and maintains a continuous, clean scalar-field representation across the imbricate thrust array and related fold belt. Specifically, the recovered horizons exhibit clear and consistent throw when crossing individual thrusts, and display geologically reasonable bending and thickness-variation trends within fold cores and the frontal wedge, in agreement with the stepwise forward propagation of deformation recorded by the syn-kinematic growth strata. Meanwhile, the model preserves stable stratigraphic continuity near the deep detachment. Overall, these results indicate that our method can produce geologically consistent implicit models in real thrust systems characterized by strong compression, non-monotonic stratigraphy, and multi-level imbrication, while exhibiting improved robustness and generalization to complex structural assemblages.

Figure \ref{fig6} (b-1) illustrates an idealized fold–and–thrust system. A series of thrust faults rise from a basal detachment and form an imbricate stack; the strata trapped between adjacent thrusts are progressively uplifted and tightened into folds. At depth, the system can be understood as slip and displacement transfer controlled by a floor thrust, whereas up-dip the geometry may evolve into a composite structure in which thrusts merge and are jointly bounded by a roof thrust, such that the structural style transitions from a frontal imbricate fan to more internal, complex repetition and shortening.

As shown in Fig. (b-3)–(b-4), our method is able to generate an implicit structural representation of an imbricate thrust array within this fold-and-thrust system.
multiple thrusts initiate from the deep floor thrust and climb ramps upward, driving systematic folding and kinematically consistent throw in the hanging-wall strata. Building on this, the model further recovers the up-dip linkage characteristic of “buried” thrusting in the hinterland, where several thrusts rejoin an upper detachment (roof thrust), forming a duplex geometry that bounds discrete horses. More importantly, the reconstructed horizons preserve continuous stratigraphic geometry and coherent topology across fault terminations, thrust convergence zones, and tightly folded cores, without obvious cross-horizon leakage or local numerical oscillations, yielding a section geometry that more closely satisfies the requirements of internally consistent structural interpretation.

The thin-skinned thrust example in Figure \ref{fig6} (c-1) corresponds to the balanced cross-section from the Bornes–Aravis region discussed by Butler et al. This thrust belt lies on the inner side of the Jura fold-and-thrust belt, where deformation is largely confined to the sedimentary cover, dominated by Mesozoic carbonates and shales. The system is regionally decoupled along a weak horizon such as Triassic evaporites, forming a basal detachment (basal detachment/sole thrust).

Our method can stably recover the detachment-controlled, forward-propagating displacement style in this thin-skinned thrust example.
As shown in Figure \ref{fig6} (c-3),(c-4), the reconstructed implicit scalar field exhibits a continuous and stable basal-detachment response at depth, enabling the overlying cover sequence to undergo systematic thrusting and folding above the detachment. This behavior is consistent with the balanced interpretation of the Bornes–Aravis thrust belt, where slip is detached within Triassic evaporites and transferred forward toward the foreland. Furthermore, the model preserves stratigraphic continuity and topological coherence near tight bends and ramp zones, producing clear and geologically reasonable throw and fold geometries as horizons cross thrust faults. 
Moreover, in intervals with pronounced mechanical stratification, the reconstruction maintains coordinated deformation among different stratigraphic units, thereby better capturing the internal geometric closure of the cover sequence and the stratification-controlled structural style.

\subsection{Data Flywheel}

Leveraging the strong out-of-distribution modeling capability of our approach, we can further construct training data with a much broader spectrum of structural styles by providing more complex and realistic structural skeletons as inputs. By subsequently assigning physically meaningful properties to the modeled geometries and performing forward modeling, we can synthesize diverse and more realistic paired datasets (e.g., seismic–RGT, seismic–impedance/velocity, and seismic–log proxies), thereby enriching both the diversity and fidelity of synthetic seismic observations. This self-reinforcing loop enables continuous expansion from implicit structural modeling to a wider range of geoscience and geophysical AI tasks, including horizon/fault interpretation, facies and geobody delineation, property inversion, uncertainty quantification, and multi-task or foundation-model pretraining.
Figure \ref{fig7} presents several examples of the physical properties used to generate the real data in this study and the corresponding seismic data.

\section{Conclusion}
This work proposes a diffusion based approach for implicit structural modeling. Although the model is trained only on stylized synthetic data of normal fault systems, it nevertheless exhibits remarkable out of distribution generalization, effectively handling a range of complex structural settings such as strike slip flower fault systems and thrust nappes, a capability that conventional methods generally lack. Compared with discriminative regression and GAN based learning frameworks, the proposed method can more stably satisfy fault and sparse horizon constraints in key structural regions, substantially reducing artifacts and local distortions while better preserving stratigraphic continuity and global topological consistency, thereby yielding more reliable and geologically consistent implicit scalar field results.
Furthermore, leveraging its robust modeling and generalization potential, we propose a "data flywheel" self-enhancing closed loop: by inputting increasingly complex and realistic structural skeletons, the model continuously generates high-quality implicit structures. Combined with property assignment and forward modeling, this process synthesizes more realistic and diverse multi-modal paired datasets, which can be applied to broader geophysical tasks and, in turn, further refine the model itself. The application of this method extends beyond traditional scientific and engineering problems like geological analysis and reservoir modeling, fostering a deep integration of geophysics and AI to drive the advancement of AI technologies in the Earth Sciences.


\bibliographystyle{IEEEtran}
\bibliography{references}
\end{document}